\documentclass[prd,preprint,linenumbers,amssymb,amsmath,aps,nofootinbib,superscriptaddress]{revtex4}
\usepackage{graphicx}
\usepackage{bm}
\usepackage{subfigure}
\usepackage{amssymb}
\usepackage{color,soul}
\usepackage{amsmath}

\begin{document}

\title{SKA sensitivity for possible radio emission from  dark matter in Omega Centauri}

\author{Guan-Sen Wang}
\affiliation{Key Laboratory of Dark Matter and Space Astronomy, Purple Mountain Observatory, Chinese Academy of Sciences, Nanjing 210023, China}
\affiliation{School of Astronomy and Space Science, University of Science and Technology of China, Hefei, Anhui 230026, China}

\author{Zhan-Fang Chen}
\affiliation{Key Laboratory of Dark Matter and Space Astronomy, Purple Mountain Observatory, Chinese Academy of Sciences, Nanjing 210023, China}
\affiliation{School of Astronomy and Space Science, University of Science and Technology of China, Hefei, Anhui 230026, China}

\author{Lei Zu \footnote{Corresponding author: zulei@pmo.ac.cn}}
\affiliation{Key Laboratory of Dark Matter and Space Astronomy, Purple Mountain Observatory, Chinese Academy of Sciences, Nanjing 210023, China}
\affiliation{School of Astronomy and Space Science, University of Science and Technology of China, Hefei, Anhui 230026, China}

\author{Hao Gong}
\affiliation{School of Mathematics and Physics, Bengbu University, Bengbu, Anhui 233030, China}

\author{Lei Feng \footnote{Corresponding author: fenglei@pmo.ac.cn}} 
\affiliation{Key Laboratory of Dark Matter and Space Astronomy, Purple Mountain Observatory, Chinese Academy of Sciences, Nanjing 210023, China}
\affiliation{School of Astronomy and Space Science, University of Science and Technology of China, Hefei, Anhui 230026, China}
\affiliation{Joint Center for Particle, Nuclear Physics and Cosmology,  Nanjing University -- Purple Mountain Observatory,  Nanjing  210093, China}


\author{Yi-Zhong Fan \footnote{Corresponding author: yzfan@pmo.ac.cn}}
\affiliation{Key Laboratory of Dark Matter and Space Astronomy, Purple Mountain Observatory, Chinese Academy of Sciences, Nanjing 210023, China}
\affiliation{School of Astronomy and Space Science, University of Science and Technology of China, Hefei, Anhui 230026, China}


\begin{abstract}

Omega Centauri, the largest known globular cluster in the Milky Way, is believed to be the remains of a dwarf galaxy's core. Giving its potential abundance of dark matter (DM), it is an attractive target for investigating the nature of this elusive substance in our local environment. Our study demonstrates that by observing Omega Centauri with the SKA for 100 hours, we can detect synchrotron radio or Inverse Compton (IC) emissions from the DM annihilation products. It enables us to constrain the cross-section of DM annihilation down to  $\sim {\rm 10^{-30}~cm^3~s^{-1}}$ for DM mass from several $\rm{GeV}$ to $\rm{100~GeV}$, which is much stronger compared with other observations. Additionally, we explore the axion, another well-motivated DM candidate, and provide stimulated decay calculations. It turns out that the sensitivity can reach  $g_{\rm{a\gamma\gamma}} \sim 10^{-10} ~\rm{GeV^{-1}}$ for $2\times 10^{-7} ~\rm{eV} < m_a < 2\times 10^{-4} ~\rm{eV}$.
\end{abstract}

\maketitle

\section{Introduction}
Modern astrophysics is still grappling with the enigmatic nature of DM. Various theories have been put forward to explain its characteristics, including Weakly-Interacting Massive Particles (WIMPs) and axions. WIMPs are a particularly well-motivated model, as they can account for the correct relic abundance of DM. As a result of WIMP annihilation. Particles such as electrons, positrons, and antiprotons can be detected by space-based telescopes~\cite{q1,q2,q3,q4}. Furthermore, these secondary cosmic ray particles would interact with their surroundings, generating radio emissions through synchrotron and IC effects~\cite{Chen_2021,Arpan_2020,Cembranos_2020,huang}.

Moreover, axions have gained popularity as a potential solution to the strong CP problem and are considered as a viable candidate for DM~\cite{Peccei_1977,Weinberg_1978,new1,new2,new3}. Axions, as well as axion-like particles, exhibit a wide range of masses, spanning from $\rm{\mu}$eV to radio frequency scales. As a result of the coupling between axions and photons, axion decay produces line radio emissions~\cite{Andrea_2018,Andrea2_2018}. Additionally, background photons can stimulate this axion decay~\cite{Andrea_2018,Andrea2_2018}. The radio signal is raised by both the synchrotron and IC of WIMPs and the decay of axions, indicating that radio observation is a potent technique for detecting DM particles.

The Square Kilometre Array (SKA) is a highly promising instrument in the field of radio astronomy~\cite{ska}, renowned for its exceptional resolution and sensitivity. It is an indispensable tool for detecting radio emissions from the annihilation products of DM particles. Extensive research has been conducted on various sources, such as the galactic center of the Milky Way, Draco (a dwarf galaxy), A2199 (a galaxy cluster), and Dragonfly 44 (a DM-rich galaxy)~\cite{Andrea_2018,Andrea2_2018,Cembranos_2020,Chen_2021,Wang_2021}. Also, DM in globular clusters has been widely discussed~\cite{1,2,3,4,5,6,7,8,9}. Omega Centauri, the largest globular
cluster in the Milky Way, is probably a remnant core of a dwarf galaxy with a considerable amount of
DM. Some previous work have revealed Omega Centauri with $J_{\rm factor} \sim 10^{22} ~\rm{GeV^2~cm^{-5}}$~\cite{Addy_2021, Javier_2021}. In addition, the gamma-ray emission from Omega Centauri has also been widely discussed as a potential outcome of WIMP annihilation \cite{Javier_2021,Profumo2017,Conrad2017}. To confirm whether the gamma-ray emission is indeed due to WIMP annihilation is also one of the motivations of this article.

In this study, we computed the adjoint radio emissions of DM in Omega Centauri, encompassing the continuum secondary synchrotron and IC radiation of Weakly Interacting Massive Particles (WIMPs) or the decay line of axions, with the sensitivity required for detection using SKA within 100 hours.

This work is organized as follows: In Section II, we present some background information, introduce the properties of Omega Centauri, and calculate the SKA sensitivity for both the continuum and line spectrum. Then, we analyze the radio emission from the secondary synchrotron and IC of WIMPs in Sections III, IV, and V before discussing the axion decay model in Section VI. Lastly, we summarize our findings and discuss the results in the final section.

\section{Omega Centauri and SKA sensitivity}

Omega Centauri is one of the most massive and brightest globular clusters in the Milky Way, located $\rm 5.4~kpc$ away from Earth with celestial coordinates
($201.7^{\circ}$, $-47.56^{\circ}$). Its high mass core and half-light radius ($\rm 7~pc$) as reported in Ref.~\cite{Baumgardt_2017,Harris_1996}, have led scientists to speculate that it is a remnant core of a captured dwarf galaxy~\cite{Bekki_2003}. Despite the numerous uncertainties of dark matter in Omega Centauri, multiple observations have strongly suggested that it could be an exceptional candidate for exploring dark matter~\cite{Bekki_2003,Brown_2019}.
Observations of the stellar kinematics in Omega Centauri reveal a component of DM estimated to be around $10^{6}~M_{\bigodot}$, which supports the assumption that it is a captured dwarf galaxy remnant~\cite{Addy_2021,Wang_2021}. Furthermore, there is evidence suggesting that the gamma-ray emission in Omega Centauri can be explained by DM annihilation~\cite{Javier_2021,Brown_2019}.

For DM density in Omega Centauri, we use the NFW profile following \cite{Wang_2021}:

\begin{eqnarray}
\rho=\frac{\rho_{\rm s}}{\frac{r}{r_{\rm s}}(1+\frac{r}{r_{\rm s}})^2}.
\label{eqnfw}
\end{eqnarray}
While the total mass of DM in Omega Centauri is consider with a high degree of uncertainty significant uncertainty. Previous studies have attempted to measure its total DM mass through velocity dispersion, different groups have gotten the results with about one order deviation, $M_{\rm{DM}} = 2.154\times10^6~M_{\bigodot}$ in~\cite{Watkins2015} and $2\times10^5~M_{\bigodot}$ in~\cite{Brown_2019}. (In~\cite{Watkins2015}, the abundance of $M_{\rm{DM}}$ is estimated based on the mass-to-light ratio and the total mass of Omega Centauri. By assuming the universality of the mass-to-light relation, we can estimate the mass of dark matter through various astronomical observations.)
In this work, we estimate the total mass of DM in Omega Centauri to be approximately $10^{6}~M_{\bigodot}$~\cite{Addy_2021,Wang_2021} and consider three sets of parameters: $\rho_{\rm s} = 27860.5~ M_{\bigodot}~\rm {pc^{-3}}$, $r_s = 1.0~\rm pc$~; $\rho_{\rm s} = 7650.59~ M_{\bigodot}~\rm {pc^{-3}}$, $r_s = 1.63~\rm pc$ and $\rho_{\rm s} = 4391.65~ M_{\bigodot}~\rm {pc^{-3}}$, $\rm r_s = 2.0~pc$~\cite{Wang_2021,Brown_2019}corresponding to $J_{\rm factor} \sim 2\times 10^{23} ~\rm{GeV^2~cm^{-5}}$. Additionally, considering the uncertainty in the total dark matter mass in Omega Centauri, we need to assess the potential impact of choosing different total dark matter masses. Due to our choice of a relatively large value for the total amount of dark matter, we explore the parameter values $\rho_{\rm s} = 765.059~M_{\bigodot}~\rm {pc^{-3}} $ and $r_s = 1.63~\rm {pc^{-3}}$, which permits us to examine the consequences on the results when exclusively reducing the total mass by an order of magnitude to approximately $10^{5}~M_{\bigodot}$ and $J_{\rm factor} \sim 2\times 10^{21}~\rm{GeV^2~cm^{-5}}.$ \textbf{Also, Ref.~\cite{Brown_2019} increases the scale radius to 7 pc, which should have a similar effect.} Using these parameters gives us a set of relatively conservative constraint results.
The velocity dispersion of DM is expected to be around $30~\rm km/s$~\cite{Wang_2021}, which is much lower than that of the Milky Way. Also, Omega Centauri is much closer to us than the dwarf galaxy. This feature makes Omega Centauri an ideal target to probe DM properties. Furthermore, the low gas density nature of the globular cluster ensures a negligible background in the radio frequency range. Despite no prior reports of radio emissions from Omega Centauri, we anticipate that the Square Kilometre Array (SKA), a highly sensitive radio telescope array with advanced energy resolution capabilities, may detect the potential radio signals.

SKA is a radio telescope under construction with high sensitivity and energy resolution~\cite{ska}. SKA1-LOW and SKA1-MID cover the frequency range from 50 MHz to 50 GHz~\cite{ska}. Previous radio telescopes, such as the Murchison Widefield Array, have achieved $\rm \sim O(mJy)$ for null-detection~\cite{Arpan_2020}. While SKA, in the near future, will be able to detect signals with $\rm {O(\mu Jy)}$ sensitivity.

The sensitivity of the SKA for specific observations can be described by the following \cite{ska}:
\begin{eqnarray}
S_{\rm {min}}=\frac{2 k_b S_D T_{\rm sys}}{\eta_s A_e (\eta_{\rm pol} t \Delta \nu)^{0.5}}.
\label{smin}
\end{eqnarray}
Here, $k_b$ represents the Boltzmann constant, and $S_D$ is a degradation factor with a value of 2 for continuum images and 2.5 for line images. The system efficiency, represented by $\eta_{\rm{s}} = 0.9$, the number of polarization states, represented by $\eta_{\rm{pol}} = 2$, the total observation time $t$, and the channel width $\Delta \nu$ are also included in the equation. The minimum detectable flux is determined by the ratio of the effective collecting area $A_e$ and the total system noise temperature $T_{\rm sys}$, which can be found in \cite{ska} for both SKA1 and SKA2 phases.

To calculate the minimum detectable flux for our analysis, we utilize the longest integration time of 100 hours, a channel width of $\Delta \nu =300 ~\rm{MHz}$ for the continuum image, and a channel width of $\Delta \nu =10^{-4} \nu$ for the line image~\cite{ska}. These values are chosen based on the velocity dispersion in Omega Centauri, which is about $\sim 30~\rm{km/s}$ \cite{Cembranos_2020, Wang_2021}.

\section{Calculation formulas}
The synchrotron and IC radiation of charged products from DM annihilation have been widely discussed. Several tools have been developed to calculate this processes, such as RX-DMFIT \cite{rx-dmfit}. In this draft, we adopt the RX-DMFIT package to calculate the flux of synchrotron and IC emission.

The equilibrium spectrum of CR $e^{+}/e^{-}$  can be calculated by following equation:
\begin{eqnarray}
\frac{\partial}{\partial t}\frac{\partial n_e}{\partial E}={\bf \bigtriangledown}[D(E,{\bf r}){\bf \bigtriangledown} \frac{\partial n_e}{\partial E}]+\frac{\partial}{\partial E}[b(E,{\bf r})\frac{\partial n_e}{\partial E}]+Q(E,{\bf r}),
\end{eqnarray}
where $n$ is the equilibrium electron density, $Q(E,{\bf r})$ is the injected electron source term, $D(E,{\bf r})$ is the diffusion coefficient and $b(E,{\bf r})$ is the energy loss term. For the steady-state solution, the left side of the above equation is equal to zero, and it can be solved numerically using the Greens function method \cite{green1,green2} in the RX-DMFIT package.

The source term here is the contribution of WIMP annihilation described by the following formula:
\begin{eqnarray}
Q(E,r)=\frac{<\sigma v>\rho_{\chi}^2(r)}{2m_{\chi}^2}\frac{dN}{dE_{\rm inj}},
\end{eqnarray}
where $dN/dE_{\rm inj}$ is the $e^{+}/e^{-}$ injection spectrum per WIMP annihilation event, and it is derived from DarkSUSY package for a given DM mass and annihilation channel. For the diffusion coefficient, we choose a simplified power law form as follows:
\begin{eqnarray}
D(E)=D_0E^\gamma,
\end{eqnarray}
where $D_{0}$ is the diffusion coefficient in the range of $\rm 10^{27} \sim 10^{29}~cm^2~s^{-1}$ \cite{d01,d02} and we choose $\gamma=1/3$ in this work though the real situation may be more complicated \cite{DAMPE2022BC}.

The energy loss term contains several contributions: synchrotron, IC, Coulomb, and bremsstrahlung losses which are described as follows
\begin{eqnarray}
b(E,{\bf r})&=&b_{\rm IC}(E)+b_{\rm Synch}(E,{\bf r})+b_{Coul}(E)+b_{\rm Brem}(E) \nonumber \\
&=&b_{\rm IC}^0E^2+b_{\rm Synch}^0B(r)^2E^2+b_{\rm Coul}^0n_e(1+log(\frac{E/m_e}{n_e})/75)\\
&+&b_{Brem}^0n_e(log(\frac{E/m_e}{n_e})+0.36), \nonumber
\end{eqnarray}
where $n_e$ is the mean number density of thermal electrons and it is about $\rm 0.05~cm^{-3}$ for GCs~\cite{47ne}. For magnetic field strength $B$, we conservatively choose $\sim 1\rm{\mu G}$ here. The energy loss coefficients are taken to be $b_{\rm syn}^0 \simeq 0.0254,~b_{\rm IC}^0 \simeq 0.25,~b_{\rm brem}^0 \simeq 1.51 $ and $ b_{\rm Coul}^0 \simeq 6.13$, all in units of $\rm 10^{-16}~GeV/s$~\cite{bloss}.

Then the integrated flux density spectrum of the synchrotron and IC can be easily calculated through the following equation
\begin{eqnarray}
S(\nu)=\int_{\Omega}d\Omega \int_{los} (j_{\rm syn}(\nu,l)+j_{\rm IC}(E_{\gamma },l))dl,
\end{eqnarray}
where
\begin{eqnarray}
j_{\rm syn}(\nu,r)=2\int_{m_e}^{M_{\chi}}dE \frac{dn_e}{dE}(E,r)P_{\rm syn}(\nu,E,r),
\end{eqnarray}
\begin{eqnarray}
j_{\rm IC}(E_{\gamma },r)=2\int_{m_e}^{M_{\chi}}dE \frac{dn_e}{dE}(E,r)P_{\rm IC}(E,E_{\gamma }).
\end{eqnarray}
Here $P_{\rm syn}$ and $P_{\rm IC}$ represent the standard synchrotron power and IC power, which refer to the number of photons of a specific frequency that an electron with a certain energy can produce. Further details about these components can be found in~\cite{syn}.

Here, we combine synchrotron and IC radiation, which provides a more accurate depiction of the scenario compared to the approach that calculates the impact of each component separately. This is particularly important when the two components have equal strengths, as their combined effect can be twice as large. These differences can result in variations in the constraints on DM.

The IC radiation has two components: IC with Cosmic Microwave Background (CMB) photons and IC with starlight. However, since the frequency of starlight is already higher than the radio band, the resulting frequency after IC is too high to be detected by SKA. Consequently, the impact of starlight photons on the measurement is negligible.

\section {THE ENERGY SPECTRUM DISTRIBUTION}

\begin{figure}[htb]
\includegraphics[width=12cm]{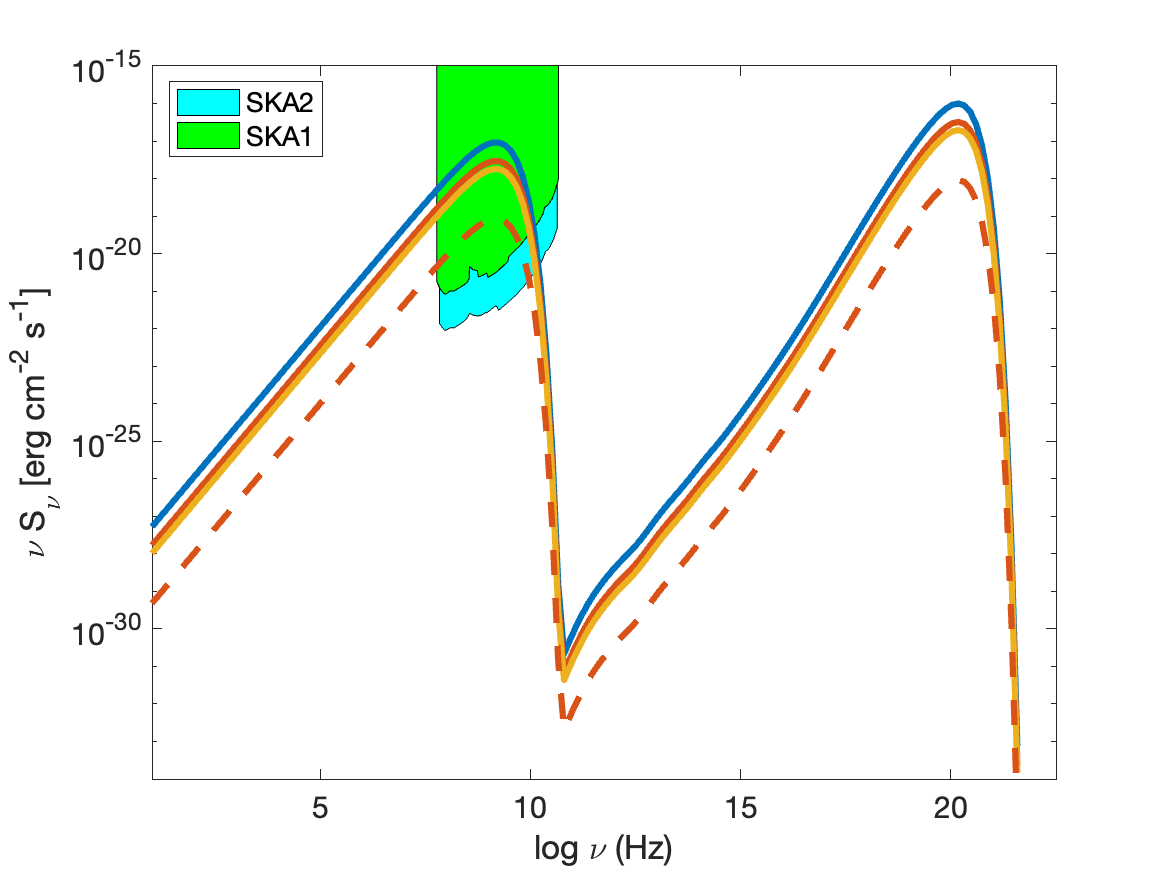}
\caption{The SED of synchrotron and IC emission for $m_\chi = 10~\rm{GeV}$, $D_0 = 10^{30}    ~\rm {cm^{2}~s^{-1}}$, $<\sigma v>=10^{-30} ~\rm{cm^{3}~s^{-1}}$, $B=1~\rm {\mu G}$ and various choice of NFW profile parameters: $\rho_{\rm s} = 7650.59~M_{\bigodot}~\rm {pc^{-3}} $, $r_s $=1.63 pc (red line), $\rho_{\rm s} = 27860.5~M_{\bigodot}~\rm {pc^{-3}} $, $r_s $=1.0 pc (blue line), $\rho_{\rm s} = 4391.65~ M_{\bigodot}~\rm {pc^{-3}} $, $r_s $=2.0 pc (yellow line), $\rho_{\rm s} = 1.61\times10^6 ~M_{\bigodot}~\rm {pc^{-3}} $ and $\rho_{\rm s} = 765.059 ~M_{\bigodot}~\rm {pc^{-3}} $, $r_s $=1.63 pc (red dashed line). The cyan (green) region represents the sensitivity of SKA phase1 (SKA phase2) with 100 hours of observation, similarly hereinafter.}
\label{SED_differentrho}
\end{figure}

To account for the uncertainty of the DM profile in Omega Centauri, we investigated the effects of different mass and NFW profile parameters as shown in Fig.~\ref{SED_differentrho}. 
Our analysis reveals that the intensity of the spectral energy distribution (SED) greatly affected by dark matter masses. A one-order smaller mass induces a two-order lower SED, as indicated by the two red lines in Fig.~\ref{SED_differentrho}. 
Moreover, when keeping the total mass constant, a smaller $r_s$ reduces the impact of diffusion and energy losses during the electron propagation process, resulting in a larger SED. However, this effect is not expected to significantly influence the final results as long as the total mass of dark matter remains constant. There remains substantial uncertainty in determining the total amount and distribution of dark matter in Omega Centauri. 
Currently, there is no way to determine the accurate value definitively, so here we choose $\rho_{\rm s}=7650.59~M_{\bigodot}~\rm{pc^{-3}}$ and $r_s=1.63$~pc as a set of typical profile for the following analysis.

\begin{figure}[htb]
\includegraphics[width=10cm]{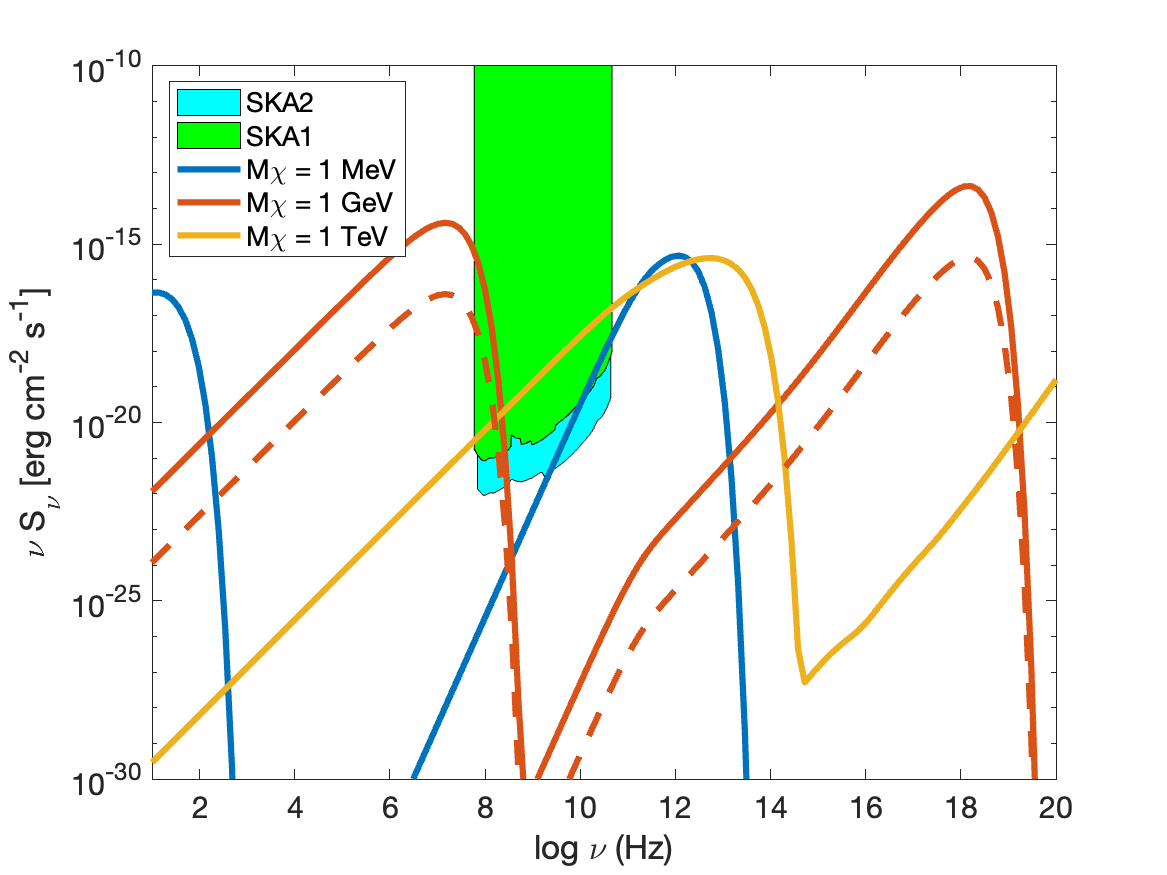}
\caption{The SED of synchrotron and IC emission for $D_0 = 10^{30}~\rm {cm^{2}~s^{-1}}$, $<\sigma v>=10^{-30} ~\rm{cm^{3}~s^{-1}}$, $B=1~\rm {\mu G}$ and various choice of DM mass: 1 MeV (blue line), 1 GeV (red line), and 1 TeV (yellow line).} 
\label{SED_differentmass}
\end{figure}

\begin{figure}[htb]
\subfigure[$B=B_{0}$]{\includegraphics[width=8.1cm]{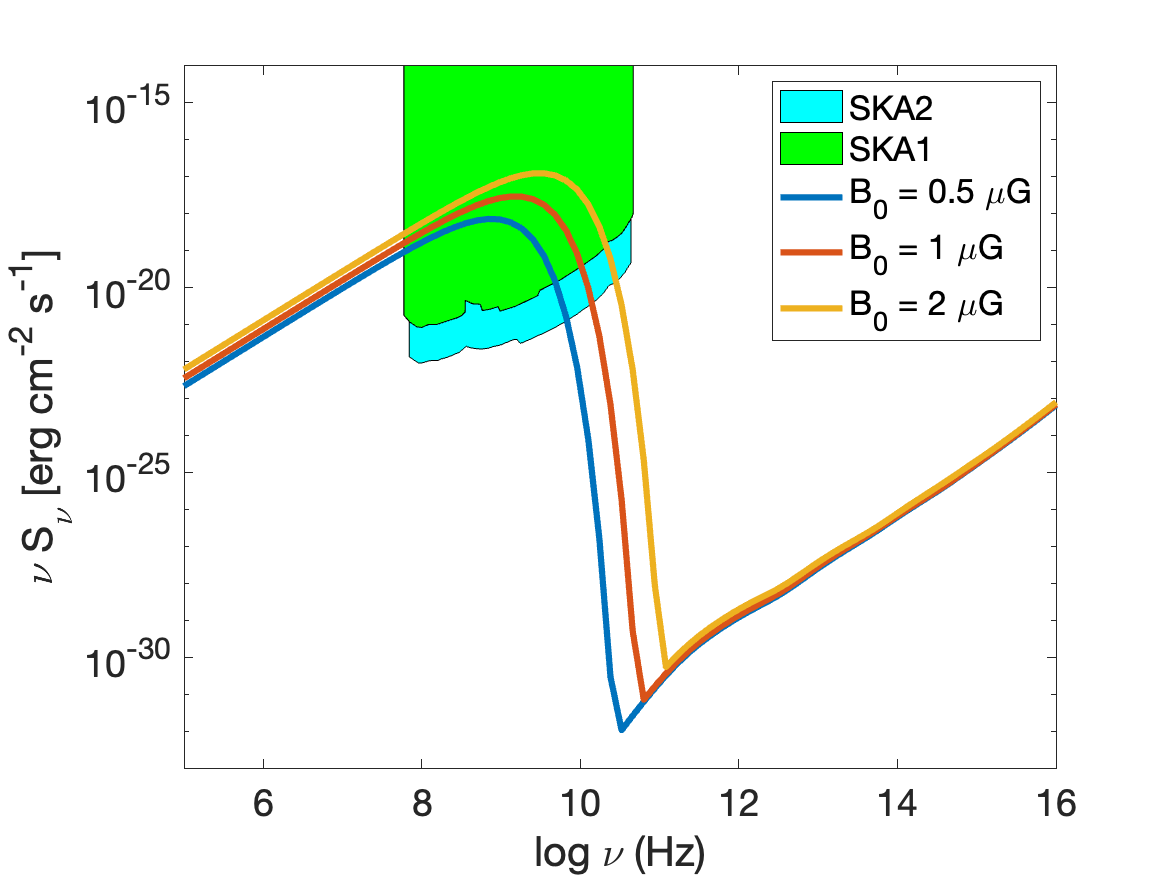}}
\subfigure[$B=B_{0}\times e^{-(r/r_{c})}$]{\includegraphics[width=8.1cm]{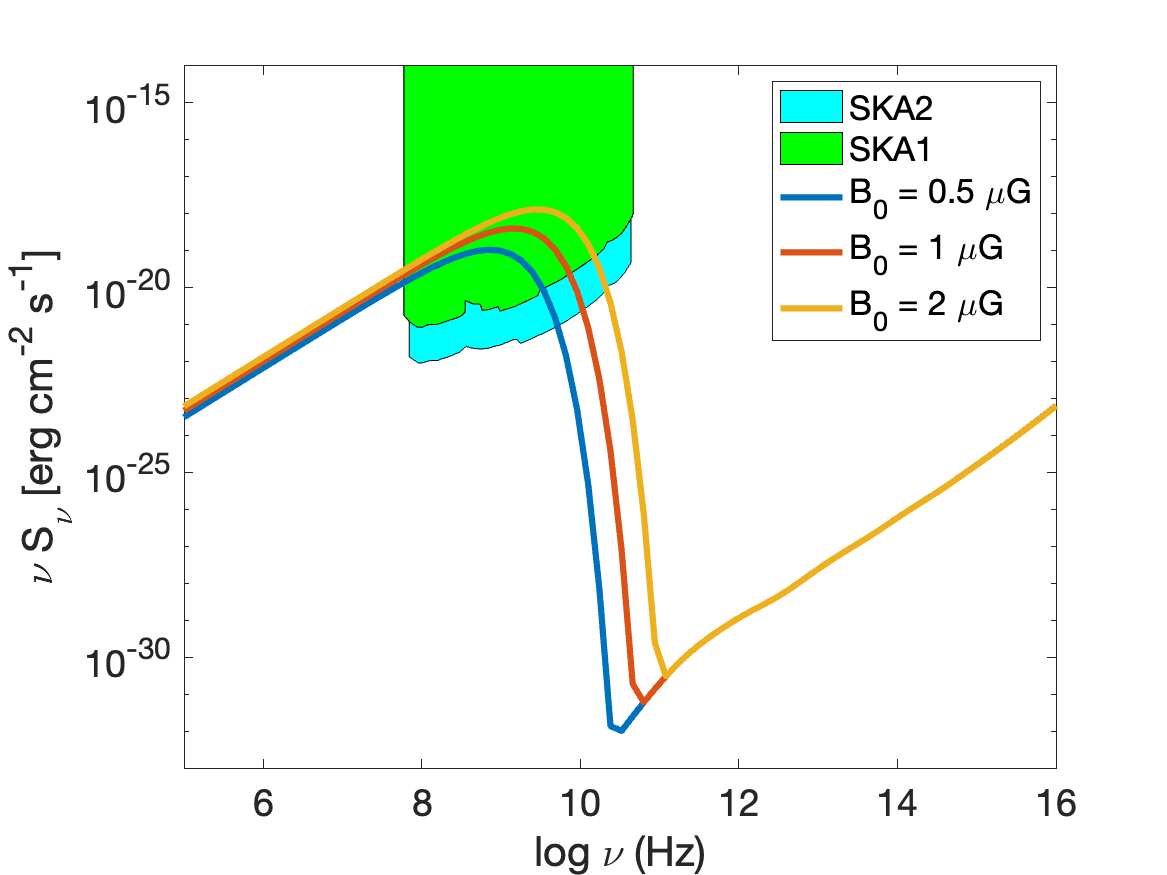}}
\caption{The SED of synchrotron and IC emission for $m_\chi =10~\rm {GeV}$, $D_0 = 10^{30}~\rm {cm^{2}~s^{-1}}$, $<\sigma v>=10^{-30} ~\rm{cm^{3}~s^{-1}}$ and various choice of magnetic field strength in constant magnetic field model (a) and exponentially decaying magnetic field model (b). }
\label{SED_differentB}
\end{figure}

The spectral energy distribution (SED) of radio emissions relies heavily on the mass of the DM particle. That is because the energy of the electron products and the corresponding photon flux is determined by the DM mass. In Fig.~\ref{SED_differentmass}, we present the results of SEDs for various $m_\chi$ with fixed diffusion coefficient, magnetic field strength and cross-section. 
For MeV DM, IC emission is the primary contributor, while synchrotron emission dominates for heavier DM particles.

\begin{figure}[htb]
\includegraphics[width=11cm]{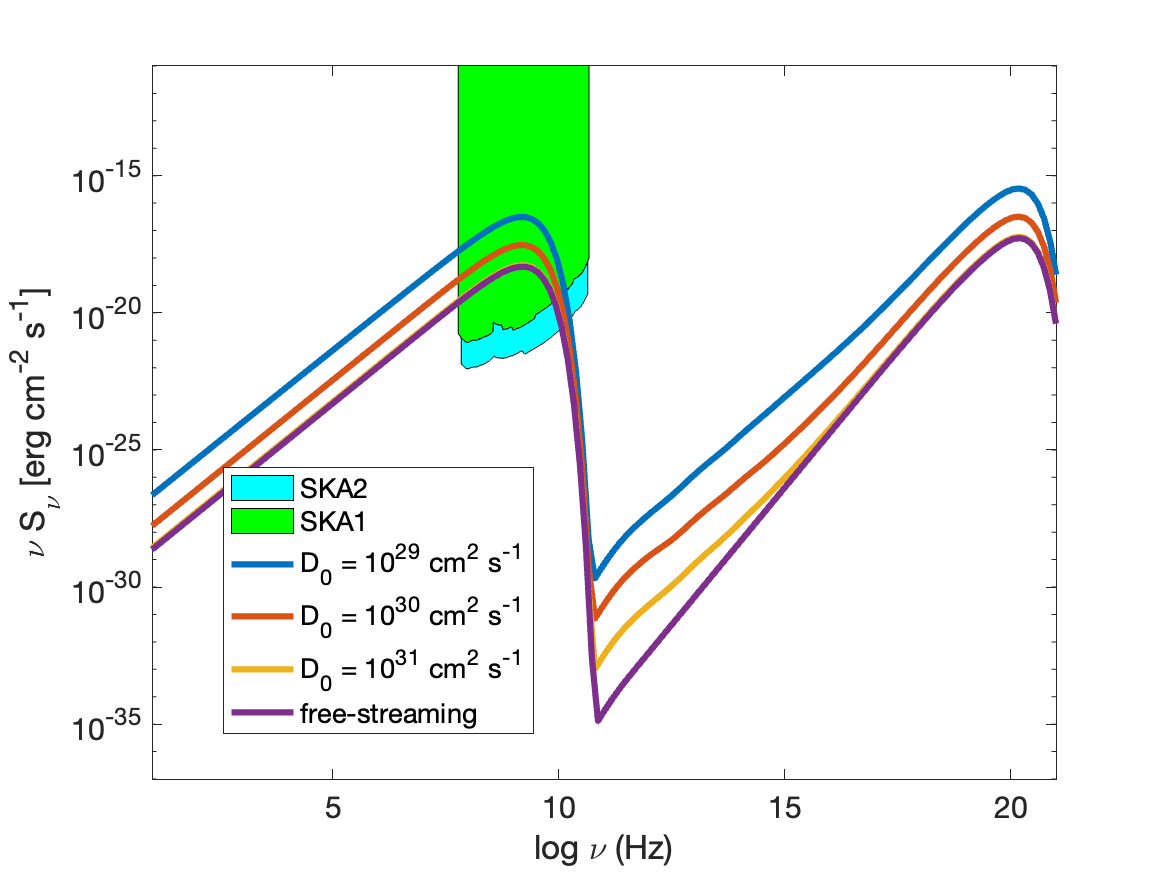}
\caption{The SED of synchrotron and IC emission for $m_\chi =10~\rm{GeV}$, $B=1~\rm {\mu G}$, $<\sigma v>=10^{-30} ~\rm{cm^{3}~s^{-1}}$ and various choice of diffusion coefficient.}
\label{SED_differentD0}
\end{figure}

The diffusion coefficient and magnetic field strength also significantly affect the final radio flux, as they impact both the charged particle propagation and photon emission processes. Their influences on SED are presented in Fig.~\ref{SED_differentB} and Fig.~\ref{SED_differentD0}. The synchrotron emission is strongly dependent on the magnetic field since the synchrotron process is produced by the electron propagation in the magnetic field, while the IC emission is not affected. To account for the uncertainty of magnetic field measurements, we explored several different magnetic field strengths as illustrated in Fig.\ref{SED_differentB}. The radio emission flux increases by nearly an order of magnitude as the magnetic field strength rises from 0.5 $\rm{\mu G}$ to 2 $\rm{\mu G}$. The influence of the diffusion coefficient is shown in Fig.~\ref{SED_differentD0}. Larger $D_0$ means that electrons are more likely to interact with gas or other materials, resulting in more significant energy losses before escaping the diffusion zone, which in turn leads to a lower flux. 
\textbf{In addition, we incorporate a free-streaming scenario to effectively bound the uncertainties in our analysis. Notably, for a diffusion coefficient of $D_0 = 10^{31}~\rm {cm^{2}~s^{-1}}$, the behavior of high-energy electrons closely resembles that of free-streaming dynamics. Consequently, the energy spectra exhibit a remarkable alignment with the free-streaming model, as illustrated in Fig.~\ref{SED_differentD0}, with only marginal discrepancies attributed to the photon spectra produced by IC process involving low-energy electrons.}

\section{THE constraints on the WIMP case}

\begin{figure}[htb]
\subfigure[~$e^+~e^-$~channel]
{\includegraphics[width=11cm]{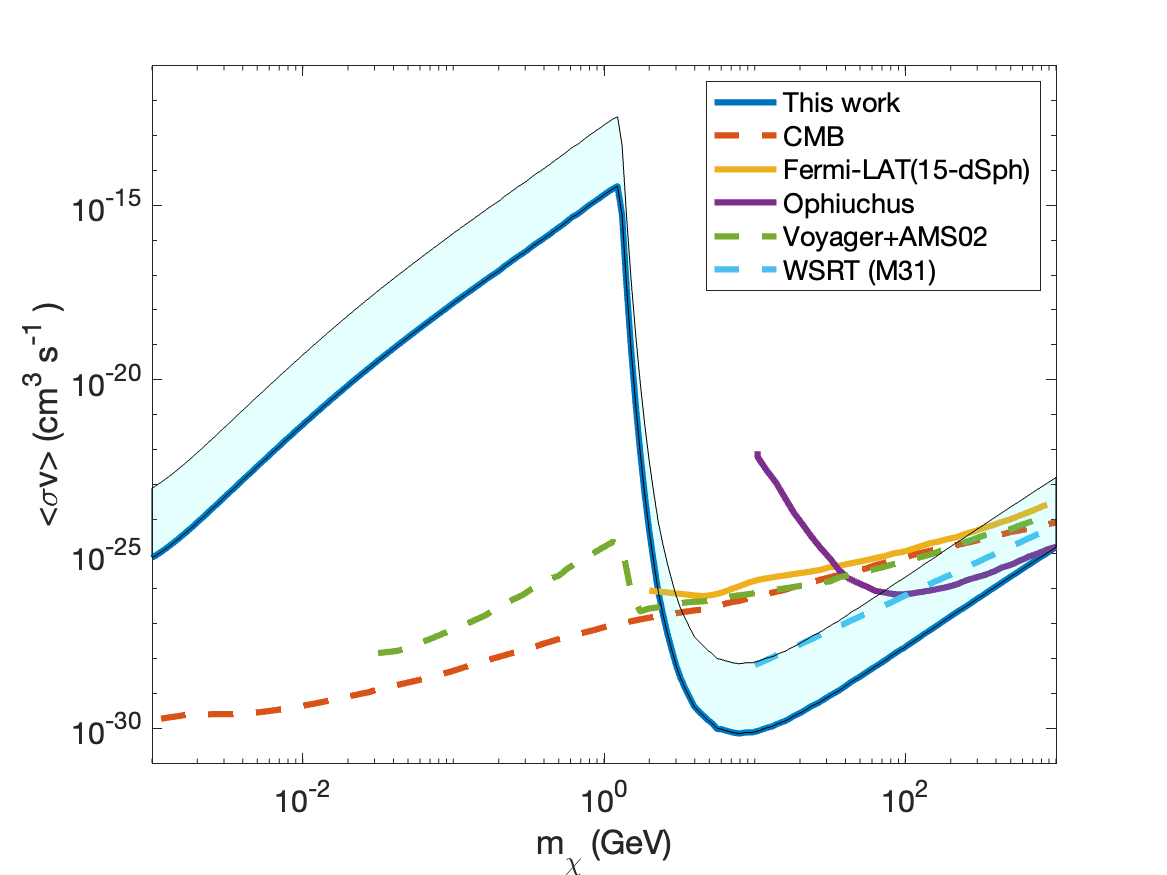}}
\subfigure[~$\mu^+~\mu^-$~channel]{\includegraphics[width=8.15cm]{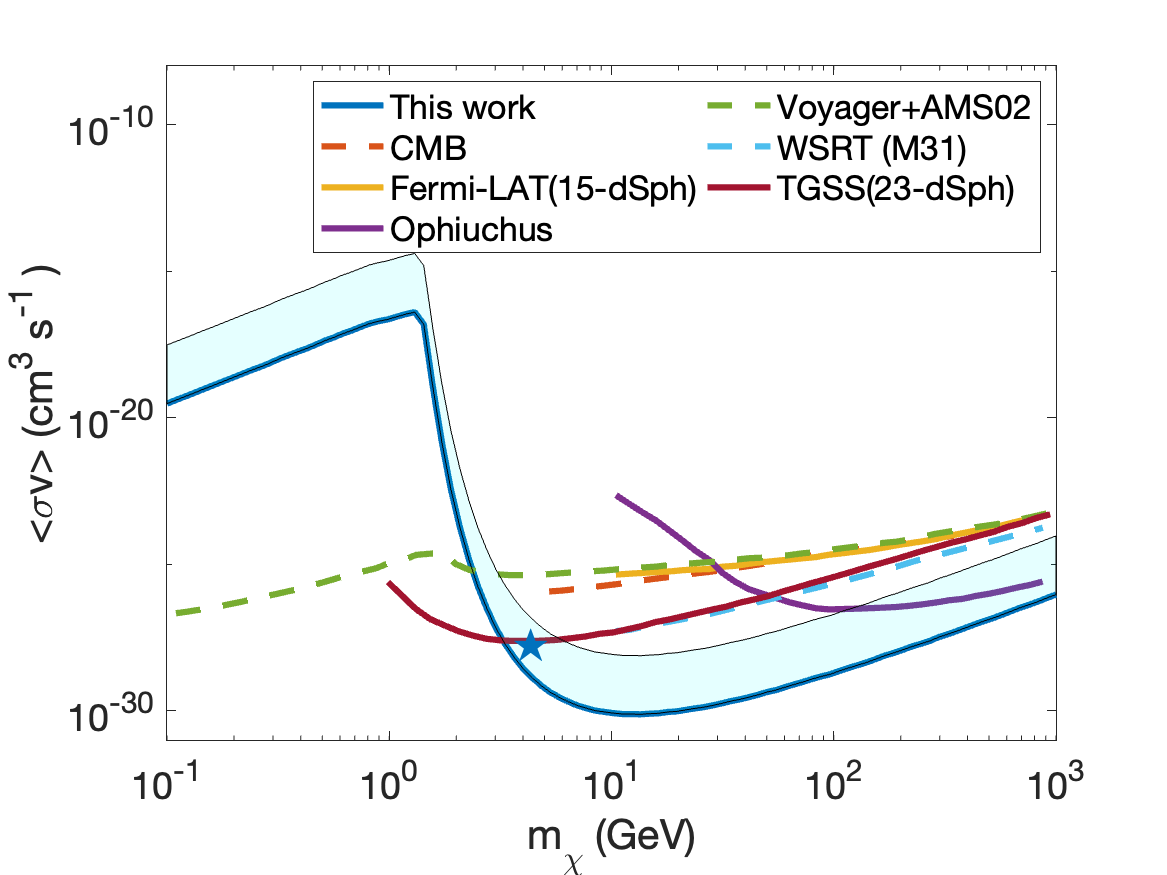}}
\subfigure[~$b\Bar{b}$~channel]
{\includegraphics[width=8.15cm]{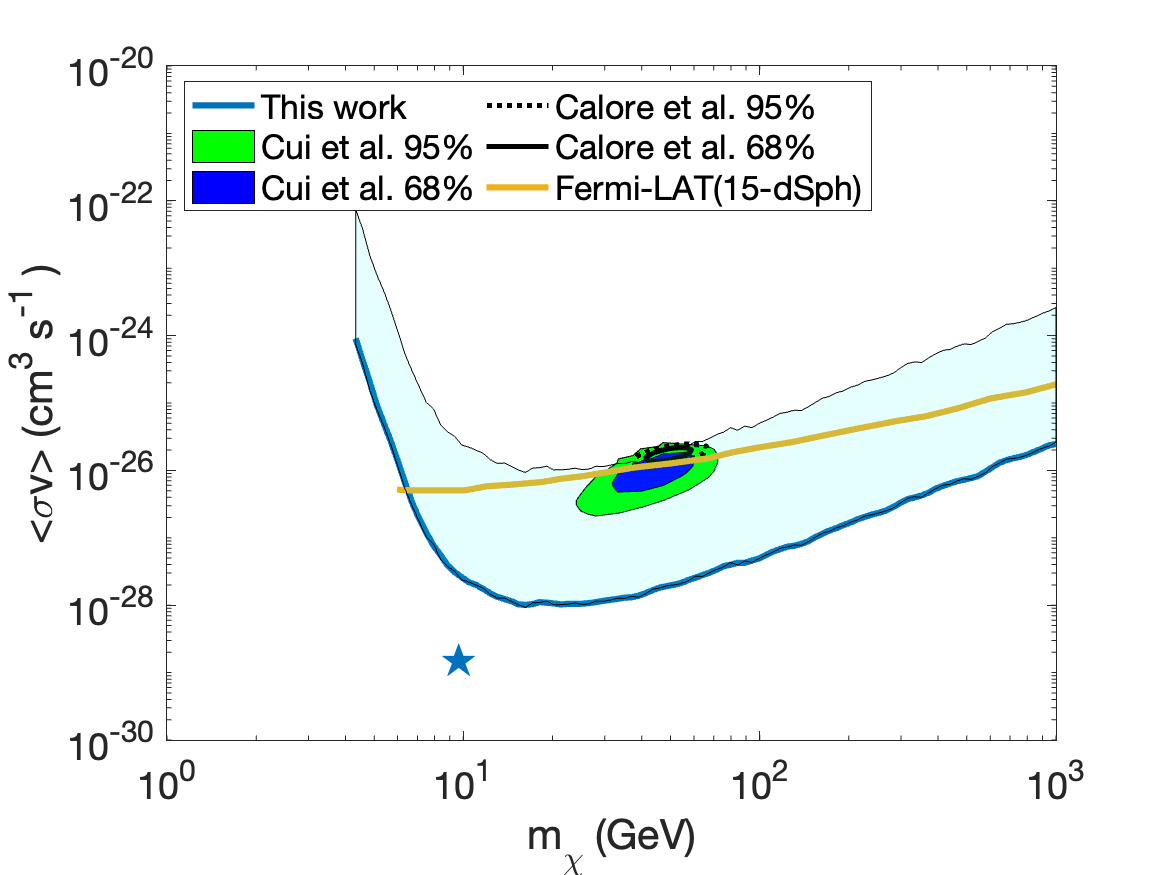}}
\caption{The exclusion limits of DM velocity-averaged annihilation cross-section for $D_0 = 10^{30}~\rm {cm^{2}~s^{-1}}$ and $B=\rm{1~\mu G}$. Comparing with other limits from different sources based on (a) ~$e^+~e^-$~channel, (b)~$\mu^+~\mu^-$~channel and $b\Bar{b}$ channel (c). The blue star representing the case where WIMPs annihilate into $\rm{\mu^+ \mu^-}$ or $b\Bar{b}$ to explain gamma-rays in Omega Centauri~\cite{Javier_2021}.}

\label{sigmav_withmu}

\end{figure}

By comparing the results of SED with SKA's sensitivity, we establish the upper limits of the DM annihilation cross-section. 
For each parameter set, we calculate the minimum cross-section required for the photons to be just detectable, i.e. the radio emission equals the SKA's sensitivity. Thus SKA offers a distinctive way of determining whether the gamma-ray emission of Omega Centauri comes from DM annihilation. The results are presented in Fig.~\ref{sigmav_withmu}.
For simplicity, we only consider the sensitivity at 1 GHz which is $2.6\times 10^{-7}~\rm{Jy}$~\cite{ska}. 

The exclusion limits of DM velocity-averaged annihilation cross-section is about $\rm 10^{-15}~cm^3~s^{-1}$ ($\rm 10^{-17}~cm^3~s^{-1}$) at sub-GeV and $\rm 10^{-30}~cm^3~s^{-1}$ at several GeV for electron-positron ($\mu^+\mu^-$) channel. For $b\Bar{b}$ channel, due to the large mass of b quark, we only have the upper limits from GeV to TeV, where radio photons are mainly produced by synchrotron emission, and the limits can reach $\rm 10^{-28}~cm^3~s^{-1}$ at 10-100 GeV range. Moreover, we use a blue shade to show the uncertainty of J-factor between $2\times10^{21}~\rm{GeV^2~cm^{-5}}$ and $2\times10^{23}~\rm{GeV^2~cm^{-5}}$. This helps demonstrate the potential impact on the final constraints when there is a lower amount of dark matter. Also, we compare our limits with other previous results from different observations: The limits from M31 by the Westerbork Synthesis Radio Telescope (WSRT)~\cite{42,43}, from 15 dwarf galaxies by Fermi LAT~\cite{44,45}, Ophiuchus by VLA ~\cite{46,47}, from 23 dwarf galaxies (TGSS)~\cite{48}, CMB data~\cite{49,50,51} and Voyager plus AMS02 data~\cite{49,50,51}. It is interesting to note that our constraints are generally stronger in the GeV-TeV range but weaker in the Sub-GeV range compared with other observations. The best-fit parameter corresponding to the gamma-ray emission of Omega Centauri~\cite{Javier_2021} is also present in Fig.~\ref{sigmav_withmu}. Here we have already calibrated the DM cross-section at $J_{\rm{factor}}=2\times10^{23}~\rm{GeV^2~cm^{-5}}$ with the result of $J_{\rm{factor}}<\sigma v>$ in Ref.~\cite{Javier_2021}.

The constraint of $b\Bar{b}$ channel is presented in  Fig.~\ref{sigmav_withmu}~(c). Unlike the $e^+e^-$ and $\mu^+\mu^-$ channel, the limit of $b\Bar{b}$ final stats are mainly from synchrotron contribution because the decay products of the b quark  have so high energy that the photons generated by the following IC process are not in the radio band. The 68\% and 95\% credible regions (shaded regions) by fitting to the antiproton data~\cite{52,53} and the 68\%, 95\% credible regions (black line) by fitting to the Galactic center GeV excess are also presented in the figure~\cite{54}. The result shows that the SKA has sufficient capability in detecting these favored parameter regions.

\begin{figure}[htb]
\subfigure[$B=B_{0}$]{\includegraphics[width=8.15cm]{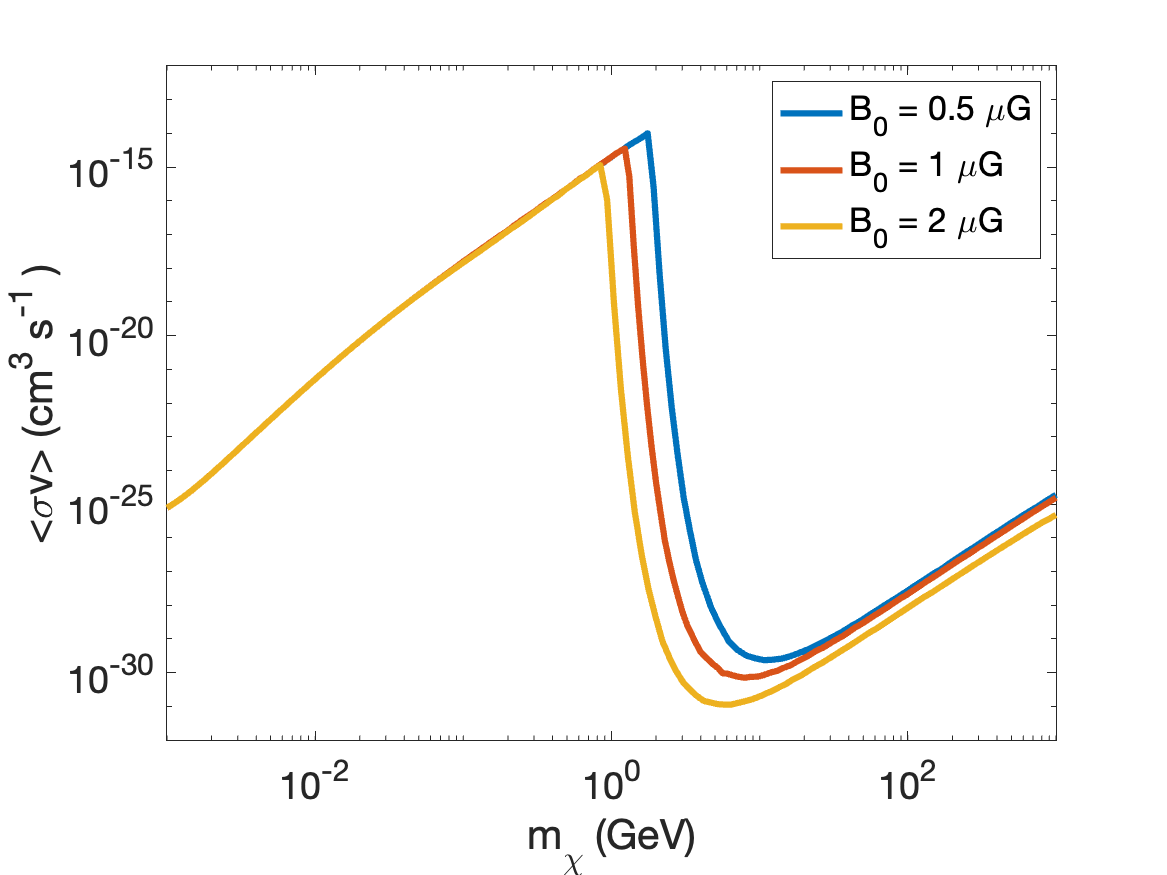}}
\subfigure[$B=B_{0}\times e^{-(r/r_{c})}$]{\includegraphics[width=8.15cm]{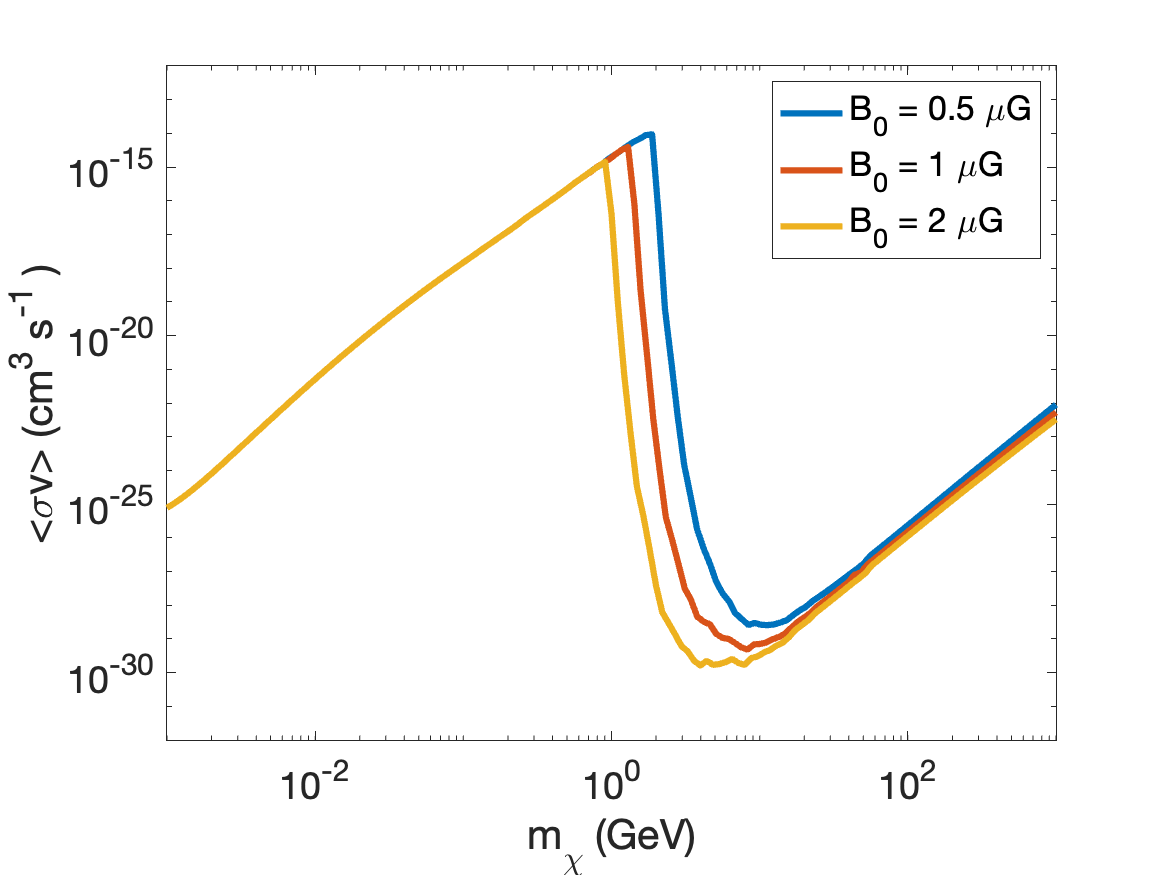}}
\caption{The exclusion limits of DM velocity-averaged annihilation cross-section on $e^+ e^-$ channel for $D_0 = 10^{30}~\rm {cm^{2}~s^{-1}}$ and different magnetic field model, including constant magnetic field (a) and exponentially decaying magnetic field (b).}
\label{sigmav_differentB}
\end{figure}
\begin{figure}[htb]
\includegraphics[width=11cm]{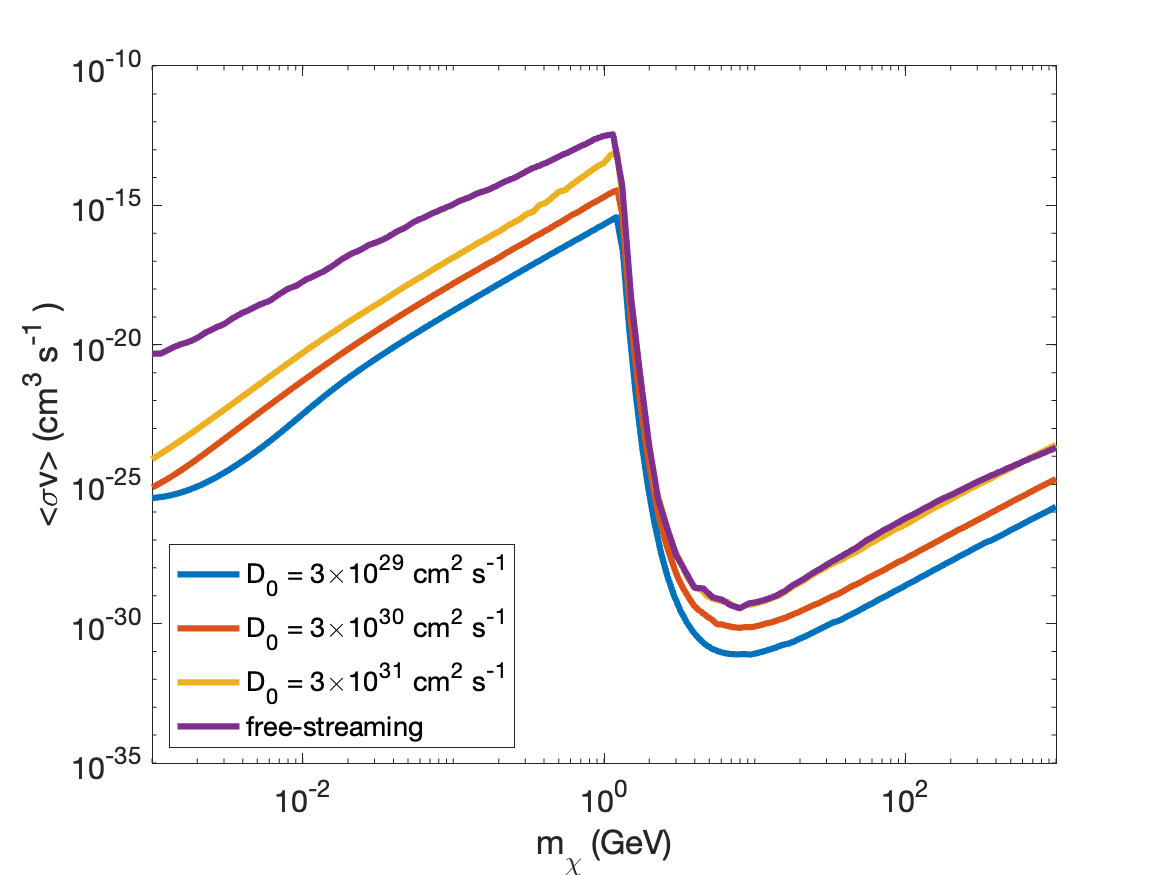}

\caption{The exclusion limits of DM velocity-averaged annihilation cross-section on $e^+ e^-$ channel for $B=\rm{1~\mu G}$ and various choice of diffusion coefficients.}

\label{sigmav_differentD0}

\end{figure}

Next, we need to discuss the impact of other parameters on the limits, here we take the $e^+ e^-$ annihilation channel as an example. As discussed earlier, the magnetic field solely affects the synchrotron radiation intensity. For high-mass WIMPs, the radio emission is predominantly from the synchrotron process, with the upper limit becoming stronger as the magnetic field increases. However, for low-mass WIMPs, the radio emission is mainly produced by IC radiation, so the upper limit remains unchanged, as depicted in Fig.~\ref{sigmav_differentB}. In contrast, the diffusion coefficient considerably impacts the upper limits of the cross-section in both synchrotron and IC cases. 
The diffusion coefficient in Omega Centauri also exhibits significant uncertainty. Currently, there is no direct observational data on the diffusion coefficient in dwarf galaxies or globular clusters. Since the exact value of the diffusion coefficient is currently unknown, and here we consider a range from $10^{29}~\rm {cm^{2}s^{-1}}$ to $10^{31}~\rm {cm^{2}~s^{-1}}$ to explore the potential impact of different diffusion coefficients. Additionally, we consider a free-streaming option to help bracket the uncertainties. As previously discussed, lower diffusion coefficients induce tight upper limits of DM cross-section as shown in Fig.~\ref{sigmav_differentD0}. \textbf{Also, we notice that for $D_0=10^{31}~\rm {cm^{2}s^{-1}}$, the propagation for high energy electrons closely approaches free-streaming and results in roughly the same upper limit.}

With a high magnetic field or low diffusion coefficient, the upper limit can even reach $10^{-31}~\rm{cm^3~s^{-1}}$ for a WIMP mass of approximately 10 GeV. Furthermore, the diffusion coefficient only affects the constraint strength, while the peak and valley position of the DM cross-section remains the same as shown in Fig.~\ref{sigmav_differentD0}.

Generally, even if accounting for uncertainties in the diffusion coefficient and magnetic field, all parameters interpreting the gamma-ray signals are expected to produce detectable radio signals for SKA.

\section{Constraints in the case of stimulated decay of axion}

\begin{figure}[htb]
\includegraphics[width=14cm]{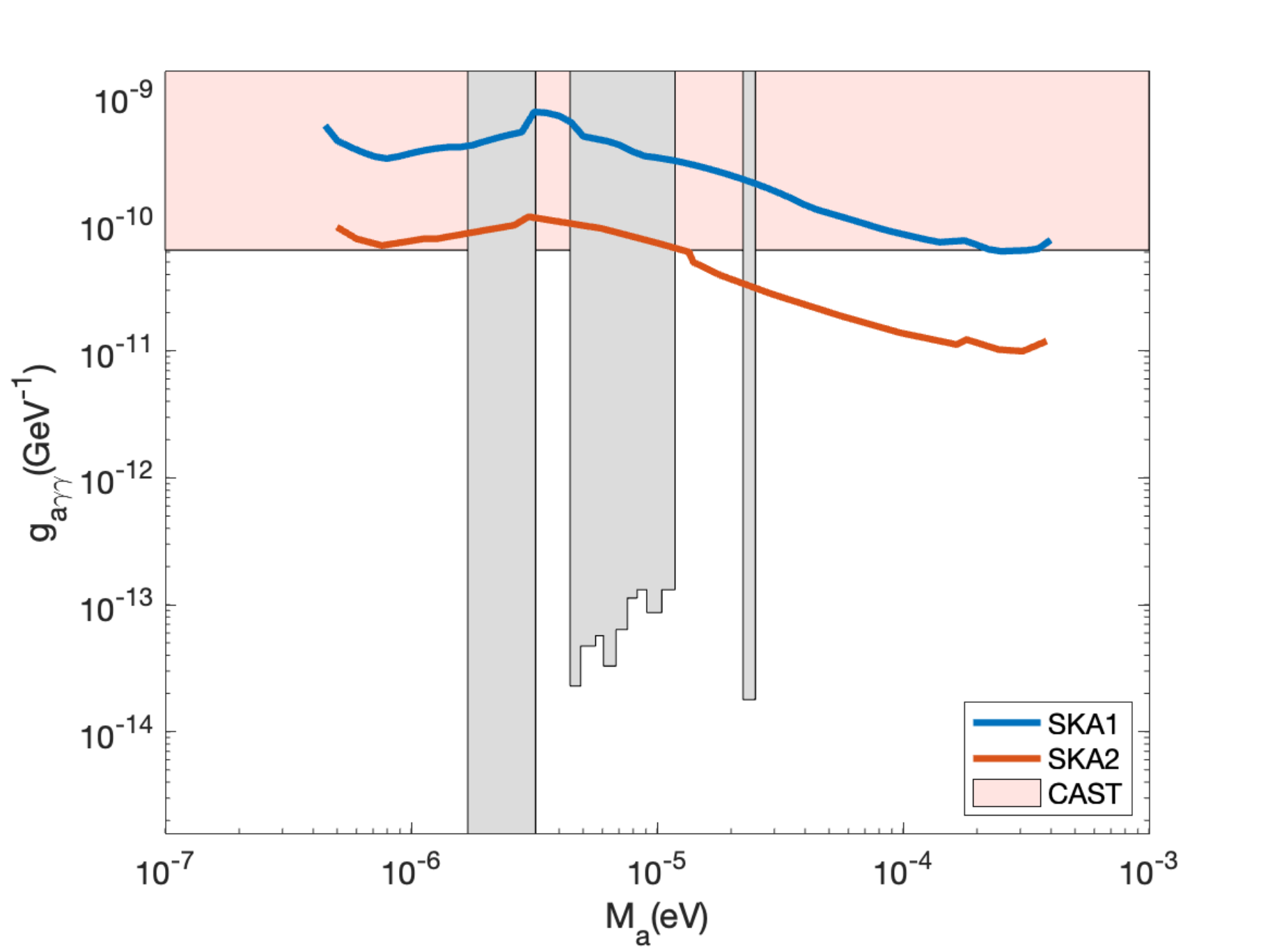}
\caption{Projected sensitivities of stimulated decay of axion for Omega Centauri with 100 hours observation in both SKA1 and SKA2 The pink regions are excluded by CAST experiments~\cite{cast}. The gray regions are excluded by experiments \cite{Hagmann_1998,Stephen_2001,Du_2018,Zhong_2018}}
\label{axion}
\end{figure}

Axions, which have a mass with the order of $\mu \rm{eV}$, are also potential DM candidates. Unlike WIMPs, they do not produce continuous synchrotron or IC radio emissions, but line structure emissions at a frequency of $\nu = \frac{m_a}{4\pi}$ through the decay of axions. The lifetime of axions in a vacuum environment is given by:
\begin{equation}
\tau_a=\frac{64\pi}{m_a^3 g_{a\gamma\gamma}^2},
\end{equation}
where $g_{a\gamma\gamma}$ is the axion-to-two-photon coupling parameter.
 
The background radio radiation at frequency $\nu = \frac{m_a}{4\pi}$ enhances the photon production rate due to the indistinguishability of photons and Bose-Einstein statistics \cite{Andrea_2018,Andrea2_2018}. The effective stimulated emission is given by:
\begin{equation}
\Gamma_{s}=\Gamma_{a}(1+2f_{\gamma}),
\end{equation}
where $\Gamma_{a}$ is the reciprocal of $\tau_{a}$, $\Gamma_{s}$ is the stimulated emission rate and $f_{\gamma}$ is the background photon occupation number. In this work, we choose a conservative estimate by considering only the CMB and extragalactic backgrounds and ignore the Milky way background. Additionally, since the Milky way background photons are mainly gathered in the galactic center and the photon energy density decreases rapidly as the radius increases~\cite{Andrea2_2018}, the galactic background is not expected to significantly impact Omega Centauri. The CMB and extragalactic backgrounds can be described by the following equation~\cite{Andrea2_2018},
\begin{eqnarray}
f_{\gamma}=f_{\rm {\gamma ,CMB}}+f_{\rm {\gamma ,exb-bkg}}.
\end{eqnarray}
The photon occupation from a blackbody spectrum is given by:
\begin{eqnarray}
f_{\rm {\gamma , bb}}=\frac{1}{e^{\frac{E_{\gamma}}{k_b T}}-1}, 
\end{eqnarray}
where $k_b$ is Boltzmann constant and T is the blackbody temperature. T=2.725 K for CMB and $T_{\rm {ext-bkg}}(\nu) \sim 1.19~(\rm{GHz}/\nu)^{2.62}~{\rm K}$ for extragalactic background. 

The flux for the line radio signal is~\cite{Andrea2_2018}:
\begin{eqnarray}
S=\frac{\Gamma_a}{4\pi \Delta \nu} \int d \Omega dl \rho_a(l,\Omega) (1+2f_\gamma),
\label{axion_sig}
\end{eqnarray}
where $\Delta \nu = (v_{disp}/c) \nu$ is the width of axion line. In this work, we use $v_{disp}=30~\rm{km/s}$ \cite{Wang_2021}.
With $D_{\rm {factor}}=\int d \Omega dl \rho_a(l,\Omega)$, we rewrite Eq.\ref{axion_sig} as:
\begin{eqnarray}
S=\frac{\Gamma_a}{4\pi \Delta \nu}  D_{\rm {factor}}(1+2f_\gamma).
\end{eqnarray}
For point-like target, $D_{\rm {factor}}$ is approximated as \cite{Lisanti_2018}:
\begin{eqnarray}
D_{\rm {factor}} \approx \frac{1}{d^{2}} \int dV \rho = \frac{M_{\rm {DM}}}{d^2},
\end{eqnarray}
where $d$ is the distance between the target and the Earth.
Thus, we can obtain the limit of the axion-photon coupling parameter with the following equation:
\begin{eqnarray}
g_{a\gamma \gamma}~>~\left[\frac{256\pi ^{2} \Delta \nu d^{2}S_{min}}{m_a^3M_{\rm DM}(1+2f_{\gamma})}\right]^{1/2}.
\end{eqnarray}

The results presented in Fig.~\ref{axion} demonstrate a continuous limit between $10^{-7}~\rm{eV} \sim 10^{-4}~\rm{eV}$, providing a significant improvement over previous studies \cite{Hagmann_1998,Stephen_2001,Du_2018,Zhong_2018}. Unlike the calculation of radio signals from axion-photon resonant conversion in the magnetospheres of magnetic white dwarf stars or neutron stars, which heavily depends on the compact stars model and magnetic field strength, our results are independent of these parameters. Furthermore, using SKA1, we obtained an upper limit of approximately $10^{-10}~\rm{GeV}^{-1}$, and with SKA2, the limit can be improved to $10^{-11}~\rm{GeV}^{-1}$, surpassing the current result from CAST. Therefore, SKA provides an excellent opportunity for the detection of low-mass axions.

\section{Conclusion}
Omega Centauri is proposed as a DM-rich globular cluster near the Milky Way. Here, we calculated the  radio emission from the DM annihilation/decay, including WIMPs and axions, and  discussed the detection capability of SKA.


The synchrotron and IC emissions from cosmic ray $e^+/e^-$ produced by WIMP annihilation provide a complementary method for detecting DM particles. We calculate the synchrotron and IC emission of WIMP annihilation in Omega Centauri
and evaluate the measurement prospects of SKA. Our calculations show that, for the dark matter parameters associated with the gamma-ray emission from Omega Centauri resulting from dark matter annihilation into muon channel, the radio emissions are detectable relative to the sensitivity of the SKA. By setting the radio emission flux from DM annihilation to be less than SKA's sensitivity at $\nu=1~\rm{GHz}$, we obtain the upper limits of DM cross-section for DM masses from MeV to TeV. The upper limits can reach $10^{-30}~\rm{cm^3s^{-1}}$ at 10 GeV, which is much stronger compared with other observations. In general, the measurement of SKA can be used to test the parameter space proposed to explain the tentative gamma-ray radiation of Omega Centauri or that involved in interpreting the Galactic GeV excess, antiproton excess and the W-boson mass anomaly \cite{Fan2022,Zhu2022} in the near future.

Next, we calculate the line emission in the radio band from the stimulated decay of axions. The projected sensitivity of SKA is shown in Fig.~\ref{axion}, and it is effective over a broad mass range of $10^{-7}-10^{-4}~\rm{eV}$. Generally speaking, SKA can probe some of the unexplored axion-photon coupling regions.

In summary, the SKA's high sensitivity allows us to investigate the untapped parameter space of DM annihilation or decay. 
With multi-frequency measurements, 
some DM candidates suggested in the literature can be effectively tested.

{\bf Acknowledgments} 
This work is supported by the National Key R\&D Program of China (Grants No. 2022YFF0503304), the National Natural Science Foundation of China (12220101003), the National Natural Science Foundation of China (Grants No. 11773075) and the Youth Innovation Promotion Association of Chinese Academy of Sciences (Grant No. 2016288). Hao Gong is supported by the Youth Talent Project of Anhui Education Department (Grant No. gxyq2019105).

\end{document}